\documentclass{ws-ijmpa}

\begin{document}

\markboth{V. Dzhunushaliev, D. Singleton, D. Dhokarh} {Effective Abelian-Higgs Theory from SU(2) gauge field theory}

%
\catchline{}{}{}{}{}
%

\title{Effective Abelian-Higgs Theory from SU(2) gauge field theory}

\author{\footnotesize  VLADIMIR DZHUNUSHALIEV
}

\address{Department of Physics and Microelectronics Engineering, Kyrgyz-Russian
Slavic University, Bishkek, Kievskaya Str. 44, 720000, Kyrgyz
Republic \\
and \\
Freie Universit\"at Berlin, Fachbereich Physik, Arnimalleee 14,
14195 Berlin, Germany \\
dzhun@hotmail.kg
}

\author{DOUGLAS SINGLETON and DANNY DHOKARH}

\address{Physics Department, CSU Fresno, 2345 East San Ramon Ave. \\
M/S 37 Fresno CA 93740-8031, USA \\
dougs@csufresno.edu ~;~ dannyx@csufresno.edu }

\maketitle

\pub{Received (Day Month Year)}{Revised (Day Month Year)}

\begin{abstract}
In the present work we  show that it is possible to arrive at a
Ginzburg-Landau (GL) like equation from pure SU(2) gauge theory.
This has a connection to the dual superconducting model for color
confinement where color flux tubes permanently bind quarks into
color neutral states. The GL Lagrangian with a spontaneous symmetry
breaking potential, has such (Nielsen-Olesen) flux tube solutions.
The spontaneous symmetry breaking requires a tachyonic mass for the
effective scalar field. Such a tachyonic mass term is obtained from
the condensation of ghost fields.

\end{abstract}

\section{Introduction}

In ref. \cite{dzhsin02a} it was shown that one could obtain London's
equation from SU(2) gauge theory using the ideas of Abelian
projection. This supports the dual superconductor picture of a
confining Yang-Mills gauge theory, since London's equation gives a
phenomenological description of the Meissner effect. In
\cite{dzhsin02a} the quantized SU(2) gauge fields split into two
phases: an ordered and  a disordered phase. The ordered phase was
the gauge field belonging to  the U(1) subgroup (the diagonal,
Abelian component of the SU(2) gauge field) and the disordered phase
was the coset SU(2)/SU(1) (the off-diagonal components of the SU(2)
gauge field). The disordered phase played a role similar to the
complex scalar field of the GL equation. In \cite{dzhsin02a} we
simply set this scalar field to a constant vacuum expectation value
{\it i.e.} we froze the equations of the SU(2)/U(1) fields making
them nondynamical degrees of freedom. In this way we obtained
London's equation. In the present work we want to ``unfreeze'' the
equations connected with the SU(2)/U(1) part, and show that it is
possible to obtain a GL-like equation. This helps strengthen the
dual superconductor picture of the confining Yang-Mills vacuum. The
GL Lagrangian density is
\begin{equation}
\label{sec1} {\mathcal L} = - \frac{1}{4}f_{\mu \nu} f^{\mu \nu} +
   \left( D_\mu \varphi^* \right) \left( D^\mu \varphi \right) -
   m^2 \left| \varphi \right|^2  - \lambda \left| \varphi \right|^4
\end{equation}
where $\varphi$ is a complex scalar field, $f_{\mu \nu} =\partial
_{\mu} a_{\nu} -\partial _{\nu} a_{\mu}$, is the field strength
tensor of the Abelian field, $a_{\mu}$, and $D_{\mu} = \partial
_{\mu} -i e a_{\mu}$ is the covariant derivative.

For the dual superconducting picture it is important that the
potential for the scalar field in (\ref{sec1}) be of the spontaneous
symmetry breaking form, and have Nielsen-Olesen flux tube solutions
\cite{no}. This means that the mass term in (\ref{sec1}) should be
tachyonic ({\it i.e.} $m^2 <0$). In this work the tachyonic mass
term is generated via the condensation of ghosts as in ref.
\cite{dudal} (see also ref. \cite{lemes}).

Strictly the dual superconducting picture of the Yang-Mills vacuum
requires a {\it dual} GL Lagrangian which gives a dual Meissner
effect with respect to the ``electric'' charges of the theory. In
the present work we arrive at the ordinary GL Lagrangian which gives
the ordinary Meissner effect with respect to the ``magnetic''
charges of the theory. However, it is simple to perform the
construction in this paper starting with the dual gauge fields and
arrive at an effective dual GL Lagrangian. A discussion of dual
symmetry and dual gauge fields for Yang-Mills theory can be found in
ref. \cite{kondo1}. An explicit example of the replacement of the
regular gauge fields with the dual gauge fields can be found in
\cite{par}.

\section{Ordered and disordered phases}

In this section we review the Abelian projection decomposition of
the  SU(2) gauge field  \cite{kondo} and necessary results from
\cite{dzhsin02a}. The SU(2) gauge fields, $\mathcal{A}_\mu =
\mathcal{A}^B_\mu T^B$, and field strength tensor,
$\mathcal{F}^B_{\mu\nu}$, can be decomposed as
\begin{eqnarray}
  \mathcal{A}_\mu & = & \mathcal{A}^B_\mu T^B = a_\mu T^a + A^m_\mu T^m ,
\label{sec2-10a}\\
  a_\mu & \in & U(1) \quad and \quad A^m_\mu \in SU(2)/U(1)
\label{sec2-10b}\\
    \mathcal{F}^B_{\mu\nu} T^B & = & \mathcal{F}^3_{\mu\nu}T^3 +
  \mathcal{F}^m_{\mu\nu}T^m
\label{sec2-10c}
\end{eqnarray}
where
\begin{eqnarray}
  \mathcal{F}_{\mu\nu} &=& f_{\mu\nu} + \Phi_{\mu\nu}
  \; \in \; U(1) ~~, ~~  \mathcal{F}^m_{\mu\nu} =  F^m_{\mu\nu} + G^m_{\mu\nu} \;
  \in  SU(2)/U(1) ,
\label{sec2-20a}\\
  f_{\mu\nu} &=& \partial_\mu a_\nu - \partial_\nu a_\mu
 ~~  and ~~
  \Phi_{\mu\nu} = g \epsilon^{3mn} A^m_\mu A^n_\nu \; \in \; U(1) ,
\label{sec2-20b} \\
  F^m_{\mu\nu}  &=&  \partial_\mu A^m_\nu - \partial_\nu A^m_\mu
 ~~  and ~~
  G^m_{\mu\nu}  =  g \epsilon^{3mn}
  \left(
  A^n_\mu a_\nu - A^n_\nu a_\mu
  \right) \; \in  \; SU(2)/U(1) ,
  \label{sec2-20c}
\end{eqnarray}
where $\epsilon^{ABC}$ are the structural constants of SU(2), $g$ is
the coupling constant, $a=3$ is the index of the Abelian subgroup,
and $m,n = 1,2$ are the indices of the coset. After performing the
above decomposition we next apply a quantization technique of
Heisenberg \cite{heisenberg} where the classical fields were
replaced by operators ($a_{\mu} \rightarrow \hat{a}_\mu$ and
$A^m_{\mu} \rightarrow \hat{A}^m_\mu$) and we then consider
expectation values of the fields. This is similar to the field
correlators approach in QCD (for review, see \cite{giacomo}) and in
ref. \cite{simonov} similar ideas were used to obtain a set of self
coupled equations for the field correlators.
\par
As in \cite{dzhsin02b} we make the following two assumptions
\begin{enumerate}
  \item
  After quantization the components $\hat A^m_\mu (x)$
  become stochastic. In mathematical terms this assumption means
  that the expectation values of the fields obey
\begin{equation}
  \left\langle A^m_\mu (x) \right\rangle = 0
  \qquad  and \qquad
  \left\langle
  A^m_\mu (x) A^n_\nu (x) \right\rangle \neq 0
  \label{sec2-60}
\end{equation}
Later we will give a specific form for the nonzero term.
   \item
  The gauge potentials $a_\mu$ and $A^m_\mu$ are not correlated, and
  $a_{\mu}$ behaves in a classical manner.
  Mathematically this means that
\begin{equation}
  \left\langle f(a_\mu) g(A^m_\nu) \right\rangle =
  \left\langle f(a_\mu) \right\rangle
  \left\langle g(A^m_\mu) \right\rangle =
  f(a_\mu)  \left\langle g(A^m_\mu) \right\rangle
\label{sec2-80}
\end{equation}
  where $f,g$ are any functions of $a _{\mu}$ and $A^m _{\mu}$ respectively.
  The classical behavior of $a _{\mu}$  results in $\left\langle f(a_\mu) \right\rangle
   \rightarrow  f(a_\mu)$.
\end{enumerate}
In distinction from ref. \cite{dzhsin02a} we will work with the
Lagrange density rather than the equations of motion of the SU(2)
gauge theory.

\section{Ginzburg - Landau Lagrangian}

We now want to show that an effective, complex, Higgs-like, scalar
field can be obtained from the SU(2)/U(1) coset part of the SU(2)
gauge theory. The self interaction of this effective scalar field is
a consequence of nonlinear terms in the original Yang-Mills
Lagrangian. The mass term for the scalar field is generated via the
condensation of ghosts fields as discussed in the following section.

Making a connection between scalar and gauge fields is not a new
idea. For example, in ref. \cite{corr} it was shown that by setting
a non-Abelian gauge field to some combination of a scalar field and
its derivatives it was possible to obtain massless $\lambda \varphi
^4$ theory. One could obtain a massive $\lambda \varphi ^4$ theory
by starting with Yang-Mills theory, but the mass term had to be
inserted by hand \cite{actor}. The final scalar field Lagrangian
that we obtain is also a massless $\lambda \varphi ^4$ theory with
the addition of a coupling to a U(1) gauge field. This is (except
for the U(1) gauge field coupling) similar to the result of
\cite{corr}. In the present paper we exchange the two gauge field of
the SU(2)/U(1) coset for a complex scalar field, whereas in refs.
\cite{corr,actor} one gauge field is exchanged for a real scalar
field. We begin by taking the expectation of the SU(2) Lagrangian
(the gauge fixing and Faddeev-Popov terms are dealt with in the next
section)
\begin{equation}
  -\frac{1}{4}\left\langle Q \left|
  \cal F^{\cal A}_{\mu \nu} \cal F^{\cal A \mu \nu}
  \right | Q \right\rangle =
  -\frac{1}{4} \left\langle
  \cal F^{\cal A}_{\mu \nu} \cal F^{\cal A \mu \nu}
  \right\rangle =
  -\frac{1}{4} \left\langle
  {\cal F}^3_{\mu \nu} {\cal F}^3_{\mu \nu} +
  {\cal F}^m_{\mu \nu} {\cal F}^m_{\mu \nu}
  \right\rangle
\label{sec3-10}
\end{equation}
First we consider the term
\begin{equation}
  \left\langle
  {\cal F}^3_{\mu \nu} {\cal F}^3_{\cal A \mu \nu}
  \right\rangle =
  \left\langle
    \left(
    f_{\mu \nu} + \Phi_{\mu \nu}
    \right)
    \left(
    f^{\mu \nu} + \Phi^{\mu \nu}
    \right)
  \right\rangle
\label{sec3-20}
\end{equation}
where the field $f_{\mu \nu}$ is in the ordered phase and $\Phi_{\mu
\nu}$ is in the disordered phase. From the second assumption of the
previous section  $a_\mu$ and $f_{\mu \nu} = \partial_\mu a_\nu -
\partial_\nu a_\mu$, behave as classical fields so that
\begin{eqnarray}
  \left\langle f_{\mu \nu} \Phi^{\mu \nu} \right\rangle  =
  f_{\mu \nu} \left\langle \Phi^{\mu \nu} \right\rangle ~,~~
  \left\langle f_{\mu \nu} f^{\mu \nu} \right\rangle & = &
  f_{\mu \nu}  f^{\mu \nu}   ~,~~
  \left\langle \Phi_{\mu \nu} \right\rangle =
  g \epsilon^{3mn} \left\langle A^m_\mu A^n_\nu \right\rangle
\label{sec3-30c}
\end{eqnarray}
For the expectation of  $A^m_\mu(y) A^n_\nu(x) $ we take the form
\begin{equation}
  \left\langle A^m_\mu(y) A^n_\nu(x) \right\rangle =
  - \delta^{mn} \eta_{\mu \nu} {\cal G}(y,x)
\label{sec3-40}
\end{equation}
${\cal G}(y,x)$ is an arbitrary function. We note that strictly
speaking this form of the expectation is an approximation since it
reduces the  degrees of freedom of the initial set of coset gauge
fields, $A^m_\mu$ (this is our purpose - to find some approximate
description of non-perturbative quantization). Nevertheless these
degrees of freedom should be quantized at the next level of
approximation. We hope that they can be quantized by perturbative
Feynman diagram techniques.

The form given in (\ref{sec3-40}) is  consistent with the color and
Lorentz indices of the left hand side. One might think to add a term
with an index structure like $\eta_{\mu \nu} \epsilon^{3 mn}$.
However, this is antisymmetric under exchange of $A^m_\mu(y)
A^n_\nu(x)$ ({\it i.e.} exchanging both Lorentz and group indices)
which is not consistent with the bosonic statistics of the gauge
fields. The quantity in (\ref{sec3-40}) is a mass dimension 2
condensate. The role of such gauge non-invariant quantities in the
Yang-Mills vacuum has been discussed by several authors
\cite{kondo1,gubarev}.

From (\ref{sec3-40}) we find
\begin{equation}
  \left\langle \Phi_{\mu \nu} \right\rangle = 0
\label{sec3-50}
\end{equation}
\begin{eqnarray}
  \left\langle
  \Phi_{\mu\nu} \Phi^{\mu\nu}
  \right\rangle &=& g^2 \Bigr(
  \left\langle A^1_\mu A^2_\nu A^{1\mu} A^{2\nu}\right\rangle +
  \left\langle A^2_\mu A^1_\nu A^{2\mu} A^{1\nu} \right\rangle  \nonumber \\
  &-&\left\langle A^2_\mu A^1_\nu A^{1\mu} A^{2\nu} \right\rangle -
  \left\langle A^1_\mu A^2_\nu A^{2\mu} A^{1\nu} \right\rangle
  \Bigr).
\label{sec3-60}
\end{eqnarray}
Approximating these quartic gauge field expectation terms as sums of
products of quadratic gauge field expectations ({\it e.g.} sums of
terms like $\delta^{mp} \delta^{nq} \eta_{\alpha\mu} \eta_{\beta\nu}
  \mathcal{G}(x_1,x_3) \mathcal{G}(x_2,x_4)$ and symmetric permutation of
  all indices and coordinates. See ref. \cite{dzhsin02b} for details) one finds
that (\ref{sec3-60}) becomes
\begin{equation}
  \left\langle
  \Phi_{\mu\nu}(x) \Phi^{\mu\nu}(x)
  \right\rangle \approx
  24 g^2 {\cal G}^2 (x,x)
\label{sec3-80}
\end{equation}
The next term from (\ref{sec3-10}) is
\begin{equation}
  \left\langle
  \mathcal{F}^m_{\mu \nu} (y)
  \mathcal {F}^{m \mu \nu} (x)
  \right\rangle \Bigr |_{y=x}
\label{sec3-90}
\end{equation}
The details of the evaluation of this expression are given in
\cite{dzhsin02b}. The evaluation of (\ref{sec3-90}) involves terms
like
\begin{equation}
  \left\langle
    \left[
    \partial_{\mu y} A^m_\nu (y)
    \right]
    \left[
    g \epsilon^{3mn} a^\mu (x)
    A^{n \nu}(x)
    \right]
  \right\rangle \Bigr |_{y=x} =
  g \epsilon^{3mn} a^\mu (x)
  \left\langle
    \left[
    \partial_{\mu y} A^m_\nu (y)
    \right]
  A^{n \nu}(x)
  \right\rangle \Bigr |_{y=x}
\label{sec3-120}
\end{equation}
which require a bit of further explanation. In eq. (\ref{sec3-40})
we excluded terms proportional to $\eta _{\mu \nu} \epsilon^{3mn}$
because of the Bose symmetry of the gauge fields. In the expression
in (\ref{sec3-120}) the gauge fields do not appear symmetrically so
such a term can be included. Thus we take the expectation of
(\ref{sec3-120}) to have the general form
\begin{equation}
  \left\langle
    \left[
    \partial_{\mu y} A^m_\alpha (y)
    \right]
  A^n_\beta
  \right\rangle =
  - \delta^{mn} \eta_{\alpha \beta} \partial_{\mu y}
  \mathcal G (y,x) - i \epsilon^{3mn} \eta_{\alpha \beta}
  \partial_{\mu y} \mathcal P (y,x)  .
\label{sec3-130}
\end{equation}
where $\mathcal P (y,x)$ is some general function. This new term
will mix gauge bosons of different colors. Using the ansatz in
(\ref{sec3-130}) and (\ref{sec3-40}) (\ref{sec3-120}) becomes
\cite{dzhsin02b}
\begin{eqnarray}
\label{sec3-185}
 \left\langle
  \mathcal F^m_{\mu \nu} \mathcal F^{m \mu \nu}
  \right\rangle &=&
-20 \Bigr[ \partial_{\mu y} \partial^\mu_x \mathcal G (y,x)
             + g^2 a_\mu (x) a^\mu (x) \mathcal G (x,x) \nonumber \\
 &-& i g a^\mu (x) \partial_{\mu x} \mathcal P^* (y,x)
+ i g a^\mu (x) \partial_{\mu x} \mathcal P (y,x) \Bigr] \Bigr
|_{y=x}
\end{eqnarray}
We now make the approximation that both functions, $\mathcal G
(y,x)$ and $\mathcal P (y,x)$, can be rewritten in terms of a single
complex scalar function as
\begin{equation}
  \mathcal G (y,x) = \mathcal P(y,x) =
  \frac{1}{5} \varphi^*(y) \varphi(x)
\label{sec3-190}
\end{equation}
Setting both $\mathcal G (y,x) $ and  $\mathcal P(y,x)$ equal to the
same product of a complex scalar field  is called the \textit{one
function approximation or ansatz}. The factor of $1/5$ is to ensure
that the kinetic term of this scalar field will have a factor of $1$
in front of it. Using (\ref{sec3-190}) in (\ref{sec3-185}) we find
\begin{equation}
  \left\langle
  \mathcal F^m_{\mu \nu} \mathcal F^{m \mu \nu}
  \right\rangle = -4\left| \partial_\mu \varphi - i ga_\mu \varphi \right|^2 \equiv
-4 \left( D_\mu \varphi^* \right) \left( D^\mu \varphi \right)  .
\label{sec3-200}
\end{equation}
Thus the total Lagrangian density is
\begin{equation}
  \left \langle {\mathcal L} \right\rangle =
  - \frac{1}{4} \left \langle
  \mathcal {F^A}_{\mu \nu} \mathcal F^{\mathcal A \mu \nu}
   \right\rangle = - \frac{1}{4}f_{\mu \nu} f^{\mu \nu} +
    \left( D_\mu \varphi^* \right) \left( D^\mu \varphi \right) -
   \frac{6 g^2}{25} \left| \varphi \right|^4
\label{sec3-210}
\end{equation}
This is the GL Lagrangian,  with a {\it massless}, effective scalar
field. This scalar field is connected with the off diagonal gauge
fields by (\ref{sec3-40}) (\ref{sec3-190}). This lack of a mass term
is a shortcoming for the effective Lagrangian  of  (\ref{sec3-210}).
Without it there is no spontaneous symmetry breaking and no
Nielsen-Olesen flux tube solutions, both which are critical to make
the connection to the dual superconducting picture of the QCD
vacuum. In the next section we show how a condensation of ghosts
fields can lead to a mass term for the effective scalar field
$\varphi$. This mass term is of the correct form ({\it i.e.}
tachyonic) to give rise to spontaneous symmetry breaking and
Nielsen-Olesen flux tube solutions.

\section{Tachyonic mass term via ghost condensation}

In ref. \cite{dudal} (see also \cite{lemes}) it was shown that a
tachyonic mass term is generated for the off-diagonal gauge fields
of a pure SU(2) Yang-Mills via a condensation of ghost and
anti-ghost fields. To the Lagrangian of (\ref{sec3-120}) one adds
gauge fixing and Faddeev-Popov terms of the form
\begin{eqnarray}
\label{sec4-20} {\mathcal L}_{GF} + {\mathcal L}_{FP} &=&
-\frac{1}{2 \alpha} (D_{\mu} ^{mn} A^{\mu n} )^2
+ i {\overline{ C}}^m D^{mp} _{\mu} D^{\mu pn} C^n \nonumber \\
&-& ig^2 \epsilon ^{mq} \epsilon ^{pn} {\overline{ C}}^m C^n A^{\mu
p} A_{\mu} ^q + \frac{\alpha}{4} g^2 \epsilon ^{mn} \epsilon ^{pq}
{\overline{ C}}^m {\overline{ C}}^n C^p C^q
\end{eqnarray}
where $\epsilon ^{mn}$ is the antisymmetric symbol for the
off-diagonal indices ($\epsilon ^{12} = -\epsilon ^{21} =1$ and
$\epsilon ^{11}=\epsilon ^{22}=0$), and $D_{\mu} ^{mn} =
\partial _{\mu} \delta ^{mn} - g \epsilon ^{mn} a_{\mu}$.
Next, replacing the quartic ghost interaction (the last term in
(\ref{sec4-20})) with an auxiliary field, $\psi$, \cite{dudal,lemes}
and applying the Coleman-Weinberg mechanism \cite{coleman} one finds
that a ghost condensation occurs with
\begin{equation}
\label{sec4-60} \langle i {\overline{ C}} ^m C^m \rangle =
-\frac{v}{16 \pi} <0
\end{equation}
since $v>0$. The third term in the Lagrangian in (\ref{sec4-20}) now
becomes
\begin{equation}
\label{sec4-70} i \epsilon ^{mq} \epsilon^{pn} {\overline{ C}}^m C^n
A^{\mu p}A_{\mu} ^q \rightarrow \frac{1}{2} \langle i {\overline{
C}} ^m C^m \rangle \langle A_{\mu} ^n A^{\mu n} \rangle =
\frac{1}{2}  \left( \frac{v}{16 \pi} \right) \left(8 {\cal G}
\right) = \frac{v}{20 \pi} \varphi ^* \varphi
\end{equation}
where (\ref{sec3-40})  and (\ref{sec3-190}) have been used. Putting
this term together with the quartic term from the effective GL
Lagrangian in (\ref{sec3-210}) we find that $\varphi$ has developed
an effective potential of the form
\begin{equation}
\label{sec4-80} V_{\varphi} = -\frac{v g^2}{20 \pi} | \varphi |^2 +
\frac{6 g^2}{25} |\varphi |^4
\end{equation}
A tachyonic mass term ($m^2 = -(v g^2)/(20 \pi) <0$)  has been
generated for the effective scalar field, $\varphi$, with the
consequence that spontaneous symmetry breaking occurs. We have
arrived at an effective {\it massive} GL-Lagrangian with the
tachyonic mass term for the $\varphi$ field  being generated by
ghost condensation. This effective GL-Lagrangian will also have
Nielsen-Olesen flux tube solutions. With this tachyonic mass term
for $\varphi$ spontaneous symmetry breaking occurs and the $U(1)$
field, $a^{\mu}$, will develop a mass of $\sqrt{ (5g^2 v)/(24 \pi
)}$. It can be shown \cite{dudal,lemes} that the ghost condensate
will not give a contribution to the mass of $a^{\mu}$, via the term
$i {\overline{ C}}^a C^a a_{\mu} a^{\mu}$.

\section{Conclusions}

In this paper we have combined several ideas (Abelian projection,
quantization methods originally proposed by Heisenberg, and some
assumptions about the forms of various expectation values of the
gauge fields) to show that one can construct an effective scalar
field within a pure SU(2) gauge field theory. The system of
effective scalar field plus the remaining Abelian field is
essentially scalar electrodynamics or the relativisitic version of
the GL Lagrangian. There is also a connection between the present
work and Cho's \cite{cho} ``magnetic symmetry'' study of the dual
Meissner effect within Yang-Mills theory. In refs. \cite{cho} a
Lagrangian similar to our eq. (\ref{sec3-210}) is obtained, with the
complex scalar field being associated with a monopole coupled to a
U(1) dual magnetic gauge boson. This may offer one possible physical
interpretation of our scalar field of eq. (\ref{sec3-190}): it may
represent some effective monopole-like field which results from the
SU(2)/U(1) coset fields. This is in accord with lattice QCD
simulations which indicate that monopole condensation plays a role
in color confinement \cite{kron}.  in ref. \cite{niemi} very similar
idea is proposed: the reduction from the SU(2) Yang-Mills
theory to a two-band dual superconductor with an interband Josephson
coupling.

The essential physics here is that one has disordered fields (the
gauge fields of the SU(2)/U(1) coset space or equivalently the
effective, complex, scalar field) which ``pushes out'' ({\it i.e.}
exhibits the Meissner effect) the ordered field (the Abelian, U(1)
field) except in the interior of the flux tubes. This is a
continuation of ref. \cite{dzhsin02a} which supports the dual
superconducting picture of the Yang-Mills vacuum for a {\it pure}
gauge field. The scalar field comes from some subset of the gauge
fields rather than being put in by hand. An interesting
question is if the procedure in this paper can be applied to the
SU(3) gauge theory of the strong interaction \cite{Dzhunushaliev:2004wh}.
In ref. \cite{par} the
SU(3) Lagrangian {\it with quarks} was studied, and using a
procedure similar to the one in the  present paper, it was found
that the dual Meissner effect did occur.

\section*{Acknowledgment}

VD is grateful Prof. H. Kleinert for invitation for the research
and Alexander von Humboldt Foundation for the support of this work.

\end{document}